\documentclass{article}
\usepackage{spconf,amsmath,graphicx}

\usepackage{color,xcolor}
\usepackage{booktabs}
\usepackage{xspace}
\usepackage{newtxmath}
\usepackage{rotating}
\usepackage{multirow}
\usepackage{url}
\usepackage{comment}
\newlength\savewidth
\usepackage[colorlinks,citecolor=citecolor, linkcolor=linkcolor, bookmarks=false]{hyperref}
\definecolor{citecolor}{HTML}{1F801F}
\definecolor{linkcolor}{HTML}{ED1C24}
\usepackage{collab}
\collabAuthor{shan}{red}{HS'Note}
\collabAuthor{yjx}{orange}{JX'Note}
\collabAuthor{wxc}{orange}{XC'Note}

\newcommand{\methodname}{MS-SENet\xspace}
\newcommand{\T}{^{\textrm T}} 

\newcommand{\vct}[1]{\boldsymbol{#1}} 
\newcommand{\mat}[1]{\boldsymbol{#1}} 
\newcommand{\etal}{\textit{et al}.\xspace}
\newcommand{\ie}{\textit{i}.\textit{e}.,\xspace}

\newcommand{\tabincell}[1]{\begin{tabular}[l]{@{}c@{}} #1\end{tabular}}
\newcommand{\addline}[1]{\multicolumn{1}{c|}{#1}}
\makeatletter
\DeclareRobustCommand{\cev}[1]{%
  {\mathpalette\do@cev{#1}}%
}
\newcommand{\do@cev}[2]{%
  \vbox{\offinterlineskip
    \sbox\z@{$\m@th#1 x$}%
    \ialign{##\cr
      \hidewidth\reflectbox{$\m@th#1\vec{}\mkern4mu$}\hidewidth\cr
      \noalign{\kern-\ht\z@}
      $\m@th#1#2$\cr
    }%
  }%
}

\title{MS-SENET: ENHANCING SPEECH EMOTION RECOGNITION THROUGH MULTI-SCALE FEATURE FUSION WITH SQUEEZE-AND-EXCITATION BLOCKS}
\name{Mengbo Li$^1$,  Yuanzhong Zheng$^1$,Dichucheng Li$^2$, Yulun Wu$^2$ Yaoxuan Wang$^1$, Haojun Fei$^{1,*}$}
\address{$^1$Qifu Technology, Shanghai, China\\
 $^2$School of Computer Science, Fudan University, Shanghai, China
}

\begin{document}
%
\maketitle
\begin{abstract}
Speech Emotion Recognition (SER) has become a growing focus of research in human-computer interaction. Spatiotemporal features play a crucial role in SER, yet current research lacks comprehensive spatiotemporal feature learning. This paper focuses on addressing this gap by proposing a novel approach. In this paper, we employ Convolutional Neural Network (CNN) with varying kernel sizes for spatial and temporal feature extraction. Additionally, we introduce Squeeze-and-Excitation (SE) modules to capture and fuse multi-scale features, facilitating effective information fusion for improved emotion recognition and a deeper understanding of the temporal evolution of speech emotion. Moreover, we employ skip-connections and Spatial Dropout (SD) layers to prevent overfitting and increase the model's depth. Our method outperforms the previous state-of-the-art method, achieving an average UAR and WAR improvement of 1.62\% and 1.32\%, respectively, across six benchmark SER datasets. Further experiments demonstrated that our method can fully extract spatiotemporal features in low-resource conditions.

\end{abstract}


\section{INTRODUCTION}

\begin{figure*}[t]
	\centering
	\includegraphics[width=1.0\textwidth]{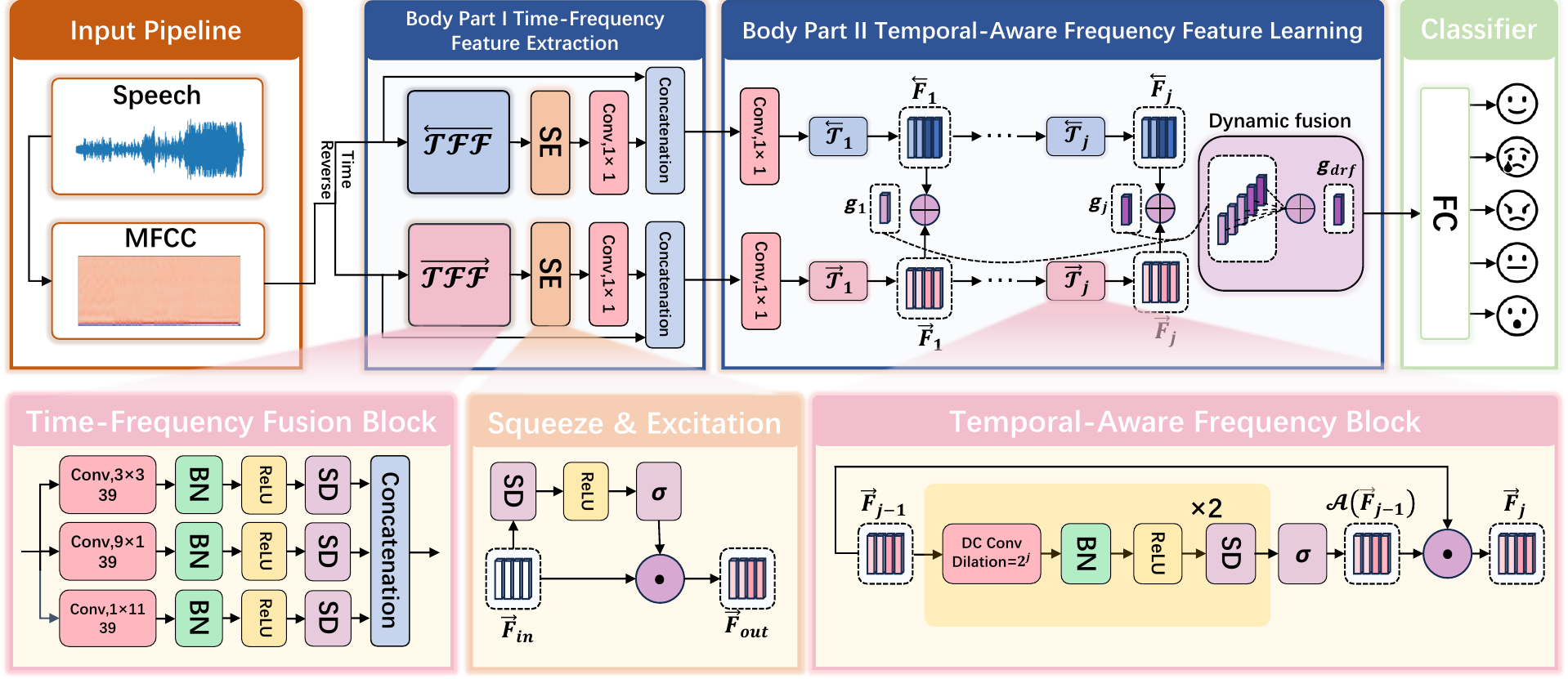}
	\caption{The MS-SENet framework for emotion feature learning consists of a Time-Frequency Fusion Block in its feature extraction component, composed of convolutional kernels in parallel paths that are eventually concatenated. Temporal-Aware Frequency Feature Learning is composed of bidirectional modules and dynamic fusion modules. $\vec{\mathcal{F}_{j}}$ represents the learned features. Note that the forward $\vec{\mathcal{T}}_{j}$, $\vec{\mathcal{TFF}}$ and backward $\cev{\mathcal{T}}_{j}$, $\cev{\mathcal{TFF}}$ have the same structure but different inputs.}
	\label{fig:architecture}
\end{figure*}
Emotion recognition plays a crucial role in human-computer interaction, especially in cases like phone customer service and voice assistants, where only speech data is available. As a result, the direct automatic identification of human emotions and emotional states from speech is growing in significance~\cite{schuller2018speech}.

Extracting features from speech has always been an important factor in recognizing speech emotion. Traditional machine learning methods~\cite{TSP_INCA,INCA_CNN}  like Support Vector Machines (SVM) mainly rely on the effective extraction of handcrafted features. In recent times, with the continuous advancement of deep learning, profound changes have been brought about in speech processing due to the ability of Deep Neural Networks (DNN) to extract intricate features directly from data.
 
Convolutional Neural Network (CNN)~\cite{RM_CNN, DBLP:conf/icpr/WenLZJ20/Capsule} are highly effective in detecting and capturing complex features in data. Yenigalla~\etal~\cite{pp} improved recognition rates by using manually crafted phoneme sequences and spectrograms.
Aftab~\etal~\cite{IEMOCAP_CNN2} extracted spectral, temporal, and spectral-temporal dependency features using three different filter sizes. Temporal neural networks like Long Short-Term Memory (LSTM)\cite{Dual_LSTM} and Temporal Convolutional Network (TCN) are widely applied in speech recognition to capture dynamic temporal changes in speech signals. 
Zhao~\etal~\cite{bilstm_ser3} extensively utilized CNN and Bi-LSTM to learn spatiotemporal features. Additionally, besides spatiotemporal features, capturing distant dependency relationships for context modeling is crucial for SER. Ye~\etal~\cite{TIM_NET} proposed a time-aware bidirectional multi-scale network called TIM-Net, achieving significant accuracy improvement in recognition.
However, these methods suffer from the following drawbacks: 1) They extracted a large number of irrelevant features without filtering, causing the discriminator to learn these features. 2) They lacked the combination of local frequency features and long-term temporal features. 3) They lacked suitable regularization mechanisms when employing deeper networks for feature extraction, increasing the risk of overfitting.

To overcome the limitations of existing methods, we propose a novel approach, and we name the network the Multi-Scale Squeeze-and-Excitation Network, termed \methodname. We employ convolutional kernels of different sizes for the initial extraction of multi-scale spatial and temporal features in speech emotion recognition. Concurrently, we introduce Squeeze-and-Excitation (SE) modules to capture and fully exploit these multi-scale features. Moreover, we employ skip-connections and Spatial Dropout (SD) layers
to prevent overfitting and increase the model's depth. 

\section{PROPOSED METHOD}
\label{METHODOLOGY}
\begin{table*}[]
\renewcommand{\arraystretch}{1.0}
\centering
\small
\caption{The overall results of different SOTA methods on 6 SER benchmark corpora. Evaluation measures are UAR(\%) / WAR(\%). The ``-"  implies the lack of this measure, and the best results are highlighted in bold.}
\begin{tabular*}{1\linewidth}{@{\extracolsep{\fill}}lcc|lcc|lcc}
\toprule[1.5pt]
\textbf{Model}    & \textbf{Year}     & \textbf{CASIA}         & \textbf{Model}   & \textbf{Year}      & \textbf{EMODB}         & \textbf{Model}   & \textbf{Year}      & \textbf{EMOVO}         \\
\midrule
TLFMRF(5-fold)~\cite{TLFMRF} & 2020        & 85.83 / 85.83           & TSP+INCA~\cite{TSP_INCA} & 2021       & 89.47 / 90.09           & ~~~~-~~~~  & ~~~~-~~~~        & ~~~~-~~~~             \\
GM-TCN~\cite{GM_TCNet} & 2022       & 90.17 / 90.17          &  Light-SERNet~\cite{IEMOCAP_CNN2}   & 2022    & 94.15 / 94.21        & TSP+INCA~\cite{TSP_INCA} & 2021       & 79.08 / 79.08           \\
CPAC~\cite{CPAC_IJCAI}  & 2022        & 92.75 / 92.75           & CPAC~\cite{CPAC_IJCAI}  & 2022         & 94.22 / 94.95          & CPAC~\cite{CPAC_IJCAI}  & 2022       & 85.40 / 85.40           \\
TIM-NET~\cite{TIM_NET} & 2023        & 94.67 / 94.67           & TIM-NET~\cite{TIM_NET} & 2023       & 95.17 / 95.70            & TIM-NET~\cite{TIM_NET} & 2023        & 92.00 / 92.00          \\\midrule
\textbf{\methodname} & 2024 & \textbf{94.95 / 95.16 } &
\textbf{\methodname} & 2024 & \textbf{96.83 / 96.40 }&
\textbf{\methodname} & 2024 & \textbf{93.31 / 93.26 }

\\\midrule\midrule
\textbf{Model}   & \textbf{Year}      & \textbf{IEMOCAP}         & \textbf{Model}    & \textbf{Year}     & \textbf{RAVDESS}       & \textbf{Model}    & \textbf{Year}     & \textbf{SAVEE}     \\
\midrule
MHA+DRN~\cite{IEMOCAP_DRN}  & 2019     & 67.40 / ~~~~-~~~~          & TSP+INCA~\cite{TSP_INCA} & 2021        & 87.43 / 87.43           &    TSP+INCA~\cite{TSP_INCA} & 2021        & 83.38 / 84.79         \\
CNN+Bi-GRU~\cite{IEMOCAP_GRU}  & 2020     & 71.72 / 70.39             & GM-TCN~\cite{GM_TCNet} & 2022       & 87.64 / 87.35           &     CPAC~\cite{CPAC_IJCAI}  & 2022         & 83.69 / 85.63           \\
Light-SERNet~\cite{IEMOCAP_CNN2}   & 2022    & 70.76 / 70.23           & CPAC~\cite{CPAC_IJCAI}  & 2022         & 88.41 / 89.03          & GM-TCN~\cite{GM_TCNet} & 2022       & 83.88 / 86.02        \\
TIM-NET~\cite{TIM_NET}  & 2023     & 72.50 / 71.65          & TIM-NET~\cite{TIM_NET} & 2023        & 91.93 / 92.08           &    TIM-NET~\cite{TIM_NET} & 2023        & 86.07 / 87.71         \\\midrule
\textbf{\methodname} & 2024 & \textbf{73.67 / 73.38 }&
\textbf{\methodname} & 2024 & \textbf{93.28 / 93.56}&
\textbf{\methodname} & 2024 & \textbf{90.00  / 90.18}\\ \bottomrule[1.5pt]

\end{tabular*}
\label{tab:SOTA}
\end{table*}

\subsection{Input Pipeline}

We use the most commonly-used Mel-Frequency Cepstral Coefficients (MFCCs) features~\cite{IEMOCAP_CNN1} as the inputs to \methodname. We set the sampling rate to match the original sampling rate of each corpus and segmented the audio signal into 50-ms frames using a Hamming window, with a 12.5-ms overlap. Following the application of a 2048-point Fast Fourier Transform (FFT) to each frame, we perform Mel-scale triangular filterbank analysis on the speech signal. Subsequently, the inverse Discrete Cosine Transform is applied to calculate the MFCC for each frame, with the first 39 coefficients selected for model training.

\subsection {Time-Frequency Feature Extraction }

We propose a novel feature fusion method, which mainly utilizes various convolutional kernels for the initial extraction of spatial and temporal features in speech emotion recognition. It employs SE modules for the selection and weighting of multi-scale features, further fusing them with the original information at multiple scales. Fig.~\ref{fig:architecture} presents the network architecture of the method in Body Part I: Time-Frequency Feature Extraction. To enhance the robustness of feature fusion, the Time-Frequency Fusion Block incorporates a SD layer. Next, we provide a detailed introduction to each component.

In the Time-Frequency Fusion Block, due to the feature extraction of multi-dimensional signals, it's customary to individually account for each dimension when calculating receptive fields. Therefore, we employed a widely-used technique, employing three parallel convolutional kernels of dimensions 9 × 1, 1 × 11, and 3 × 3 to extract the spectral, temporal, and spectral-temporal dependencies in MFCCs.

\begin{figure}[t]
	\centering
	\includegraphics[width=0.7\linewidth]{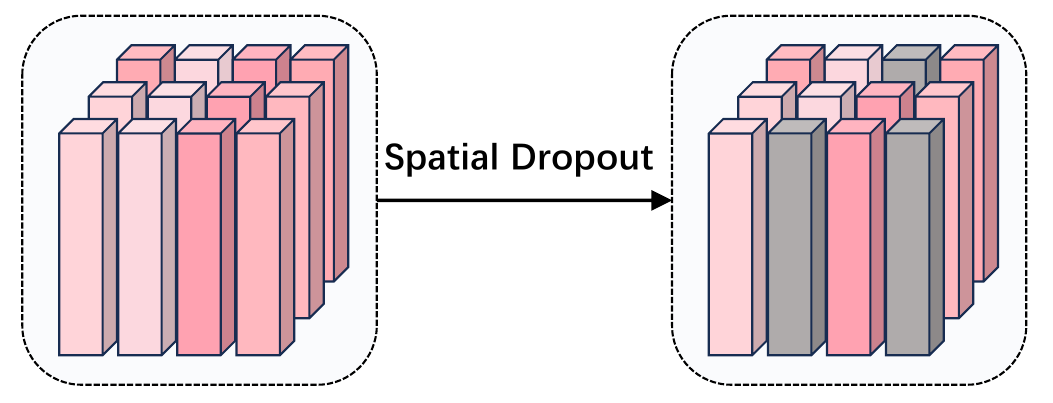}
	\caption{Illustration of Spatial Dropout.}
	\label{fig:sd}
\end{figure}

In the subsequent stages, we opted not to employ pooling layers and instead utilized SD layers. As shown in Fig.~\ref{fig:sd}, SD layer can randomly sets all values in a specific dimension to zero. This decision was mainly driven by the following reasons: 1) Pooling layers, during downsampling, may result in the loss of certain original feature information, potentially leading to the omission of emotion-relevant details and thus impacting emotion recognition performance. 2) Emotional information might not solely depend on localized acoustic features but could also correlate with the distribution and temporal aspects of the entire signal. Incorporating spatial dropout layers can maintain positional invariance to some extent, enhancing the model's adaptability and capacity to better capture emotion-relevant information. Subsequent experiments also indicated the superiority of SD layers over pooling layers, with specific details presented in the results section. Finally, the features extracted from each pathway are concatenated and input into the SE layer.

\noindent\textbf{Squeeze-and-Excitation Module.}\quad We introduce the classic SE module into the model to perform weighted adjustments on feature channels, enhancing the effectiveness of multi-scale features. The SE module aims to adaptively learn relationships between feature channels to better capture critical information. Specifically, we utilize the SE module to apply importance weighting to multi-scale features, strengthening those relevant to emotion recognition. The formula for the SE module is as follows:
\begin{align}
\mathbf{s} = \mathbf{F}_{ex}(\mathbf{z}, \mathbf{W}) = \sigma(g(\mathbf{z}, \mathbf{W}))= \sigma(g(\mathbf{W}_2 \delta(\mathbf{W}_1\mathbf{z}))),
\end{align}
In this formula, $\delta$ represents the ReLU function, $\sigma$ represents the sigmoid function, $\mathbf{W}_1\in\mathbb{R}^{\frac{C}{r}\times C}$ and $\mathbf{W}_2\in\mathbb{R}^{C\times\frac{C}{r}}$ are two fully connected matrices, and $\mathbf{r}$ is the number of intermediate hidden nodes.
After obtaining the gated unit $\mathbf{s}$, the final output $\tilde{\mathbf{X}}$ can be computed by:
\begin{align}
\tilde{x}_c = \mathbf{F}_{scale}(\mathbf{u}_c, s_c) = s_c\cdot\mathbf{u}_c.
\end{align}

\subsection{Temporal-Aware Frequency Feature Learning }
In the classification network, we employ TIM-Net~\cite{TIM_NET}, a method that learns remote emotional dependencies from both forward and backward directions while capturing multi-scale features at the frame level. The detailed network architecture of TIM-Net is illustrated in Bodypart II of Fig.~\ref{fig:architecture}. It is a bidirectional network comprising different time-aware blocks, and as the forward structure is the same as the backward, we will briefly introduce the forward structure of this network.

TIM-Net employs TAB($\mathcal{T}$) for the selection of temporal features. Each of $\mathcal{T}$  consists of two sub-blocks and a sigmoid function $\sigma(\cdot)$ to learn temporal attention maps $\mathcal{A}$, so as to generate time-aware features $\mat{F}$ through input and $\mathcal{A}$. For the two identical sub-blocks of the $j$-th $\mathcal{T}_j$, each sub-block starts by adding a DC Conv with the exponentially increasing dilated rate $2^{j-1}$ and causal constraint. 
The DC Conv is then followed by a batch normalization, a ReLU function, and a spatial dropout. The input $\vec{\mat{F}}_{j}$ for each $\vec{\mathcal{T}}_{j+1}$ is derived from the previous TAB, as shown in the Eq.~\eqref{eq:forward}:
\begin{align}
    \vec{\mat{F}}_{j+1}= \mathcal{A}(\vec{\mat{F}}_{j})\odot \vec{\mat{F}}_{j},\label{eq:forward}
\end{align}
where $\vec{\mat{F}}_{0}$ comes from the output of the first $1\times 1$ Conv layer.

Then, the bidirectional features are integrated and processed through a dynamic fusion structure as follows:
\begin{align}
    \vct{g}_j&=\mathcal{G}(\vec{\mat{F}}_{j}+\cev{\mat{F}}_{j}), \label{equ:GAP_feat}
\end{align}
\begin{align}
    \vct{g}_\text{drf}=\sum\nolimits_{j=1}^{n}w_j\vct{g}_j,  \label{equ:dynamic_rf}
\end{align}
where the global temporal pooling operation $\mathcal{G}$ takes an average over temporal dimension. $\mat{w}_\text{j}$ reprensents the items of  $\mat{w}_\text{drf} = [w_1,w_2,\ldots,w_{n}]\T$ , and they are trainable parameters. 


\section{EXPERIMENTS}
\label{EXPERIMENTS}

\subsection{Experimental Setup}
\noindent\textbf{Datasets.}\quad
In the experiments, six datasets in different languages are employed, including the Institute of Automation of Chinese Academy of Sciences (CASIA)~\cite{CASIA}, Berlin Emotional dataset (EmoDB)~\cite{EMODB}, Italian language dataset (EMOVO)~\cite{EMOVO}, interactive emotional dyadic motion capture database (IEMOCAP)~\cite{IEMOCAP}, Surrey Audio-Visual Expressed Emotion dataset (SAVEE)~\cite{SAVEE}, and Ryerson Audio-Visual dataset of Emotional Speech and Song (RAVDESS)~\cite{SAVEE}. The six datasets include 1200, 535, 588, 5531, 480 and 1440 data.

\noindent\textbf{Implementation details.}\quad
We implement the proposed models by using the Tensorflow deep learning framework. For the acoustic data, 39-dimensional MFCCs are extracted from the Librosa toolbox~\cite{librosa}.
For the $\mathcal{TFF}$, the dropout rate is 0.2. For the $j$-th TAB $\mathcal{T}_j$, there are 39 kernels of size 2 in Conv layers, the dropout rate is 0.1, and the dilated rate is $2^{j-1}$.  We set the number of TAB $n$ in both directions to 10. 
For fair comparisons with the SOTA approaches, we perform 10-fold cross-validation (CV) as well as previous works~\cite{IEMOCAP_CNN2,CPAC_IJCAI,IEMOCAP_DRN} in experiments. 

\noindent\textbf{Evaluation metrics.}\quad Due to the class imbalance, we use two widely-used metrics, Weighted Average Recall (WAR) (\ie accuracy) and Unweighted Average Recall (UAR), to evaluate the performance of each method. WAR uses the class probabilities to balance the recall metric of different classes while UAR treats each class equally.

\subsection{Results and Analysis}
\noindent\textbf{Comparison with SOTA methods.}\quad
To demonstrate the effectiveness of this approach on each corpus, we employed a 10-fold CV strategy. Table~\ref{tab:SOTA} presents the results across the six datasets, illustrating that our method consistently outperforms all previous models significantly. Prior to this, TIM-Net had already achieved the State-Of-The-Art (SOTA) on all six datasets. Compared to the original TIM-Net, our feature fusion approach improved the UA and WA of the network by 1.31\% and 1.61\%, respectively. Furthermore, it can be observed that \methodname maintains excellent emotion recognition capabilities even in cases with a higher number of emotion categories and lower data volume, such as in the SAVEE dataset.

\begin{figure}[t]
	\centering
	\includegraphics[width=1.0\linewidth]{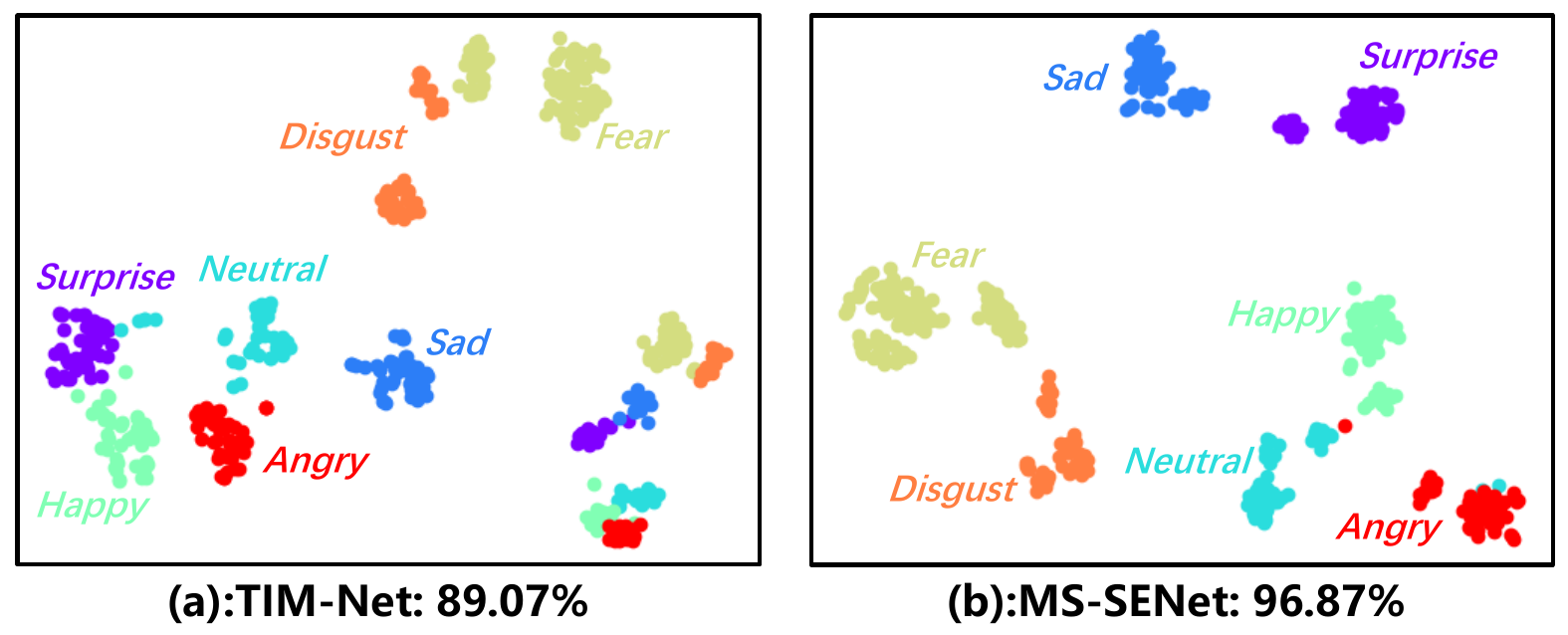}
	\caption{t-SNE visualizations of features learned from SOTA method TIM-Net and \methodname. The score denotes WAR.}
	\label{fig:tsne}
\end{figure}

\noindent\textbf{Visualization of learned affective representation.}\quad
In order to explore the influence of \methodname on representation learning, we visualized the representations learned by \methodname and TIM-Net\cite{TIM_NET} using the t-SNE technique~\cite{T-SNE} in Fig.~\ref{fig:tsne}
For a fair comparison, we use the same 8:2 hold-out validation on SAVEE corpus for the two methods and visualized the representations of the same test data after an identical training phase.
Although TIM-Net also focuses on multi-scale and temporal modeling, Figure.~\ref{fig:tsne}(a) shows a heavy overlap between \emph{Angry} and \emph{Happy}. Conversely, Fig.~\ref{fig:tsne}(b) suggests that distinct representations cluster with well-defined classification boundaries. The results confirm that \methodname offers more class-discriminative representations through effective multi-scale feature extraction and learning, contributing to superior performance.

\subsection{Ablation Study}
We conduct ablation studies on all the corpus datasets, including the following variations of \methodname: \textbf{TIM}: the \methodname is replaced with TIM-Net; \textbf{w/o SD}: the SD layer is replaced with the AVGPooling layer.; \textbf{w/o PC}: three parallel convolutional kernels are configured as 3×3.; \textbf{w/o SE}: the SE module has been removed. The results of ablation studies are shown in Table~\ref{tab:ablation}. Specific details are provided in the subsequent sections.

\noindent\textbf{Impact of SE Moudle:}\quad
 We assessed the impact of the SE module. Compared to not including the SE module, the concurrent use of SE increased WA and UA on the six corpus by 1.28\% and 0.96\%, respectively. To ensure a fair comparison, the network's other structures remained unchanged. The results indicate that the SE module can effectively learn relationships between feature channels, thereby capturing key emotional features more effectively.

\noindent\textbf{Impact of spatial dropout:}\quad
In this task, compared to using AVGPooling, using the SD layer for regularization in $\mathcal{}{TFF}$module increased WA and UA on the six corpus by 3.72\% and 2.93\%, respectively. However, when using average pooling layers, the accuracy was even lower than the baseline model. This is because average pooling layers led to the loss of original features, resulting in a decrease in model accuracy.

\noindent\textbf{Impact of parallel convolutional kernels size:}\quad
Here, we evaluated the size of parallel convolutional kernels on six corpora. Under various sizes of convolutional kernels, WA and UA increased by 0.72\% and 0.96\%. This indicates that this approach can better extract multi-scale features, thereby improving recognition accuracy.

\begin{table}[]
\renewcommand{\arraystretch}{1.0}
\small
\centering
\caption{The average performance of ablation studies and \methodname under 10-fold CV on all six corpora. The ``w/o" means removing the component from \methodname}
\begin{tabular*}{1\linewidth}{@{\extracolsep{\fill}}lccccc}
\toprule[1.5pt]
\addline{\textbf{Method}}  & \textbf{\tabincell{TIM} } & \textbf{\tabincell{w/o SD}} & \textbf{\tabincell{w/o PC}} & \textbf{\tabincell{w/o SE}} & \textbf{\tabincell{MS-SE}}\\\midrule
\addline{\tabincell{\boldmath $\mathrm{UAR}_\mathrm{avg}$\\\boldmath $\mathrm{WAR}_\mathrm{avg}$}} & \tabincell{88.76\\88.97}             & \tabincell{86.62\\87.39}               & \tabincell{89.35\\89.87}              & \tabincell{89.06\\89.36}                   & \tabincell{\textbf{90.34}\\\textbf{90.32}}\\
\bottomrule[1.5pt]

\end{tabular*}
\label{tab:ablation}
\end{table}


\section{CONCLUSIONS}
\label{CONCLUSIONS}
In this paper, we present a novel feature fusion approach, termed \methodname, which is designed for the extraction, selection, and weighting of spatial and temporal multi-scale features. It subsequently integrates these features with the original information through multi-scale fusion to better represent the features. Our experimental results demonstrate that the fusion and selection of multi-scale features play a significant role in improving SER tasks. Additional ablation studies further clarify the process of this feature fusion method, providing a new avenue for a deeper understanding of the temporal evolution of speech emotion\footnote{Code: https://github.com/MengboLi/MS-SENet}.

\vfill\pagebreak

\end{document}